\documentclass[aps,pra,twocolumn,superscriptaddress,nofootinbib]{revtex4-1}
\usepackage{amsfonts}
\usepackage{psfrag}
\usepackage{graphicx}
\usepackage{cancel}
\usepackage{amsmath}
\usepackage{natbib}
\usepackage{xr}
\usepackage{hyperref}
\usepackage{verbatim}
\usepackage{epstopdf}
\usepackage{subcaption}
\usepackage{multirow}

\usepackage{booktabs,dcolumn,caption}

\newcolumntype{d}[1]{D{.}{.}{#1}}

\begin{document}

\title{Cooling Fermions in an Optical Lattice by Adiabatic Demagnetization}
\author{Anthony E. Mirasola}
\email{tony.mirasola@gmail.com}
\affiliation{Department of Physics and Astronomy, Rice University, Houston, Texas 77005, USA}
\author{Michael L. Wall}
\altaffiliation[Current address: ]{The Johns Hopkins University Applied Physics
Laboratory, Laurel, MD 20723, USA}
\affiliation{JILA, NIST, and University of Colorado, Boulder, Colorado 80309-0440, USA}
\author{Kaden R. A. Hazzard}
\affiliation{Department of Physics and Astronomy, Rice University, Houston, Texas 77005, USA}

\begin{abstract}
The Fermi-Hubbard model describes ultracold fermions in an optical lattice and exhibits antiferromagnetic long-ranged order below the N\'{e}el temperature. However, reaching this temperature in the lab has remained an elusive goal. In other atomic systems, such as trapped ions, low temperatures have been successfully obtained by adiabatic demagnetization, in which a strong effective magnetic field is applied to a spin-polarized system, and the magnetic field is adiabatically reduced to zero. Unfortunately, applying this approach to the Fermi-Hubbard model encounters a fundamental obstacle: the $SU(2)$ symmetry introduces many level crossings that prevent the system from reaching the ground state, even in principle. However, by breaking the $SU(2)$ symmetry with a spin-dependent tunneling, we show that adiabatic demagnetization can achieve low temperature states. Using density matrix renormalization group (DMRG) calculations in one dimension, we numerically find that demagnetization protocols successfully reach low temperature states of a spin-anisotropic Hubbard model, and we discuss how to optimize this protocol for experimental viability. By subsequently ramping spin-dependent tunnelings to spin-independent tunnelings, we expect that our protocol can be employed to produce low-temperature states of the Fermi-Hubbard Model.
\end{abstract}

\maketitle

\section{Introduction}
Ultracold fermions in an optical lattice can be quantitatively described by the Fermi-Hubbard model \cite{esslingerReview}, a central model of solid state physics that can describe Mott insulators, antiferromagnetism, and potentially high-temperature superconductivity\cite{Scalapino2006,background}. Despite substantial progress in cooling fermionic gases into the Mott insulating phase \cite{EsslingerDifficult,lowneel,background,HuletMott,EsslingerMott,Schneider}, the antiferromagnetic phase has remained out of reach due to the low transition temperature (N\'{e}el temperature). Reaching the antiferromagnetic phase requires temperatures $T\sim J^2/U$, the superexchange energy, where $J$ is the tunneling and $U$ is the on-site interaction \cite{TL07}.

Many cooling schemes have been proposed for Fermi-Hubbard and other atomic systems. These include evaporative cooling \cite{evaporate}, entropy expulsion \cite{TL09,expulsion1,expulsion2}, entropy localization \cite{localize,scalettarloc}, disorder induced cooling \cite{disorder}, and conformal cooling \cite{conformal}. 
Progress has been made in a number of recent experiments \cite{Hulet,Greiner,Zwierlein,BlochAFM,Kohl,Bloch3site}. 
In Ref. \cite{Hulet}, antiferromagnetic correlations were observed at a temperature 1.4 times the transition temperature, a regime where high-temperature series expansions are not valid.
The longest range antiferromagnetic correlations were observed in Ref. \cite{Greiner} with an exponential correlation length of $\xi\sim 8$ sites in a square lattice, consistent with a temperature of $T\sim 0.2J$. 

Nevertheless, these methods have not reached regimes with  a temperature well below superexchange and a correlation length longer than a handful of sites. Even the strongest cooling methods, as demonstrated in Ref. \cite{Greiner}, reach only $T\sim 0.2 J$ and $\xi \sim 8$. Moreover, these methods rely on potential shaping techniques that are difficult to transplant to higher-dimensional systems. The temperatures are limited by various sources of heating and atom decay in the system. For example, the heating due to spontaneous emission can be on the order of $1k_B$ per second \cite{emission,heating}. Other heating processes include three-body collisions \cite{background}. In order to reach long-ranged antiferromagnetic order, the initial entropy needs to be removed substantially faster than the rate at which these heating mechanisms introduce entropy.

In many atomic and solid state systems systems, cooling has been achieved through adiabatic demagnetization \cite{IsingQuench,synthIsing,ketterle,heisenberg,heisenbergth,TL07,spinmix,crystal,dwave,etacond,inhomog,simulator,solidstatebook}. Spins begin aligned with an extremely strong external magnetic field, so that the system is in the ground state. As the field is slowly ramped to zero, the system remains in the ground state, ending close to the desired ground state of the model. Such a protocol has recently been successful in cooling ions obeying Ising dynamics to a temperature displaying antiferromagnetism \cite{monroeIsing}. 

However, as explained below, such a protocol fails for the Fermi-Hubbard model because the spin-polarized ground state of a strong external field is not adiabatically connected to the ground state of the Fermi-Hubbard model,

\begin{equation}
H_\mathrm{FH}=
- J \sum_{\left < i,j \right >, \sigma} \left (c^\dagger _{j \sigma} c^{\phantom\dagger} _{i \sigma} + \mathrm{h.c.} \right )
+ U \sum_i n_{i\uparrow} n_{i\downarrow}
\label{eq:FH}
\end{equation}
Here $\sigma \in \{ \uparrow, \downarrow\}$, $\sum_{\langle i,j \rangle}$ sums over nearest neighbors, $J$ is the tunneling energy, and $U$ is the onsite interaction energy. The fermionic annihilation operator $c_{i\sigma} ^{\phantom\dagger}$ destroys a particle of spin $\sigma$ on site $i$, and $n_{i\sigma} = c_{i\sigma} ^\dagger c_{i\sigma} ^{\phantom \dagger}$ is the corresponding number operator. 

The reason that adiabatic demagnetization fails for the Fermi-Hubbard model is its global $SU(2)$ spin symmetry. An external field breaks this symmetry down to the $U(1)$ symmetry associated with rotation about the magnetization direction, but even with this reduced symmetry, the magnetization along the magnetic field direction is a conserved quantity. Hence, the initial state, with maximal magnetization along this direction, is not adiabatically connected to the zero-magnetization ground state of interest. Rather, as the magnetic field is ramped to zero, levels cross and the initial magnetized state remains an eigenstate, merely becoming a highly excited one.

We show that by breaking the $SU(2)$ symmetry of the Fermi-Hubbard model by making the tunnelings spin-dependent, adiabatic preparation may be used to prepare low temperature antiferromagnetic states. Such spin-dependent tunnelings can be implemented in experiment in a few alternative ways, as described in Sec. \ref{sec:experiment}. With spin-dependent tunnelings, the initial ground state of the strong external field is adiabatically connected to a low energy antiferromagnetic state. We demonstrate  this explicitly in one dimensional systems numerically using the density matrix renormalization group (DMRG) method \cite{DMRG,MPS}, but we expect the idea to work more generally, both for higher dimensions and for preparing other low-temperature states.

Using DMRG calculations~\cite{wall2012out,OSMPS}, we compute the one-dimensional ground state phase diagram for the spin-anisotropic model, and we provide and analyze adiabatic demagnetization protocols to reach its ground state. We demonstrate in 1D that these protocols successfully reach low-temperature states that exhibit long-range, antiferromagnetic spin correlations for the system at half-filling, and that exhibit observables that are close to their ground state values. The model parameters necessary to implement this procedure are readily achievable in current experiments.

In Sec. \ref{sec:model}, the spin-anisotropic Hubbard model is discussed. Sec. \ref{sec:ramps} shows that low-temperature states can be obtained via ramps with modest timescales. In Sec. \ref{sec:experiment}, we discuss the experimental viability of our protocol. Sec. \ref{sec:Conclusion} concludes.

\section{Spin-Anisotropic Hubbard Model\label{sec:model}}

We study a Fermi-Hubbard model with spin-dependent tunnelings on a one-dimensional lattice of $L$ sites,
\begin{equation}
H_\mathrm{A}=
- \sum_{\left < i,j \right >, \sigma}
J_\sigma \left (c^\dagger _{j \sigma} c^{\phantom\dagger} _{i \sigma} + \mathrm{h.c.} \right )
+ U \sum_i n_{i\uparrow} n_{i\downarrow}
\label{eq:anisotropic}
\end{equation}
Here $J_\sigma$ is a spin-dependent tunneling energy. We consider positive $J_\sigma$ and $U$. Throughout, we will focus on the model at half-filling, i.e. with one atom per site, although many of the ideas and results are more general. The usual Fermi-Hubbard model is recovered when $J_\uparrow=J_\downarrow$. 

When $J_\uparrow \neq J_\downarrow$, $H_A$  lacks the $SU(2)$ rotation symmetry that impeded adiabatic demagnetization for the usual Fermi-Hubbard model. One can imagine a two-step protocol to reach low temperatures of the usual Fermi-Hubbard model: (1) First adiabatic demagnetization is used to produce a low-temperature ground state of $H_A$. (2) Then the tunneling rates are adiabatically changed to $J_\uparrow = J_\downarrow$. 

In this paper we restrict ourselves to studying the first step of this procedure. Besides its interest as a step in producing low-temperature ground states of the usual Hubbard model, the low-temperature states of the spin-anisotropic $H_A$ may be of interest in their own right. The model is challenging to solve and strongly correlated, similar to the usual Hubbard model, and we find that its ground states have similar antiferromagnetic magnetic properties to the usual Hubbard model but with stronger long-range order, as we demonstrate below.

To characterize the magnetic order, we consider spin-spin correlations $\left < S^\alpha_i S^\alpha_j \right >$ between sites $i,j$ where $\alpha\in\{x,y,z\}$ and 
\begin{align}
S_i ^x = &\frac{1}{2} \left( c^\dagger _{i,\downarrow}c^{\phantom\dagger} _{i,\uparrow} +c^\dagger _{i,\uparrow}c^{\phantom\dagger} _{i,\downarrow} \right), \\
S_i ^y = i &\frac{1}{2}\left(c^\dagger _{i,\downarrow} c^{\phantom\dagger} _{i,\uparrow} - c^\dagger _{i,\uparrow} c^{\phantom\dagger} _{i,\downarrow}\right), \\
S_i ^z = &\frac{1}{2} \left( n_{i,\uparrow}-n_{i,\downarrow}\right).
\end{align} 
Although for the ordinary Hubbard model $\left < S_i^\alpha S_j^\alpha \right >$ is independent of $\alpha$ due to the $SU(2)$ symmetry, this is not true for $H_A$. However, $\left < S^x_i S^x_j \right > = \left <S^y_i S^y_j \right >$ due to the remaining $U(1)$ symmetry of $H_A$.  We also consider the on-site number fluctuations, 
\begin{equation}
\sigma_n^2=\langle n^2\rangle - \langle n\rangle ^2,
\end{equation}
and the energy.

As $J_\sigma$ become large, we approach the limit of noninteracting particles, and long-range correlations should go to zero. In the limit $J_\sigma \ll U$, the model $H_A$ approaches an XXZ model,
\begin{equation}
H_I = \sum_{\langle i,j\rangle} K_z S_i ^z S_j ^z +K_\perp \left( S_i ^x S_j ^x + S_i ^y S_j ^y \right),
\end{equation}
where $K_z= \left( J_\uparrow ^2 + J_\downarrow ^2 \right) /2U$ and $K_\perp = J_\uparrow J_\downarrow /U$ \cite{demler}. When one of the $J_\sigma$ goes to zero, we recover an Ising model. Thus in this $J_\uparrow/J_\downarrow\rightarrow 0$ (or $J_\downarrow/J_\uparrow\rightarrow 0$) limit, we expect $\langle S^z _i S^z _j\rangle$ correlations to dominate over $x$ and $y$ spin-correlations.

\begin{figure*}[ht]
{\centering
\begin{subfigure}[]{0.49\textwidth}
\centering
\includegraphics[width=.98\linewidth]{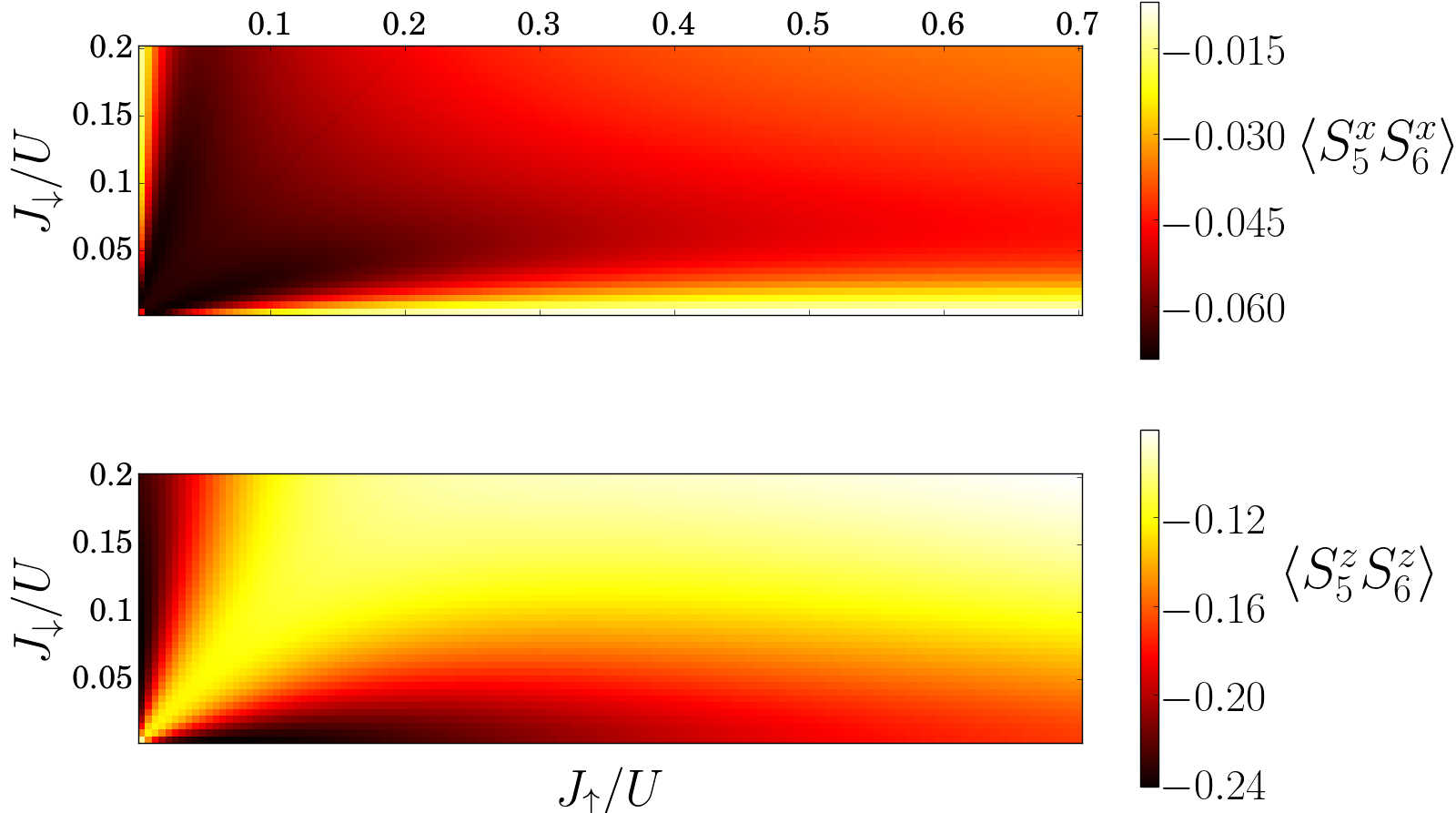}
\end{subfigure}
~
\begin{subfigure}[]{0.49\textwidth}
\centering
\includegraphics[width=.98\linewidth]{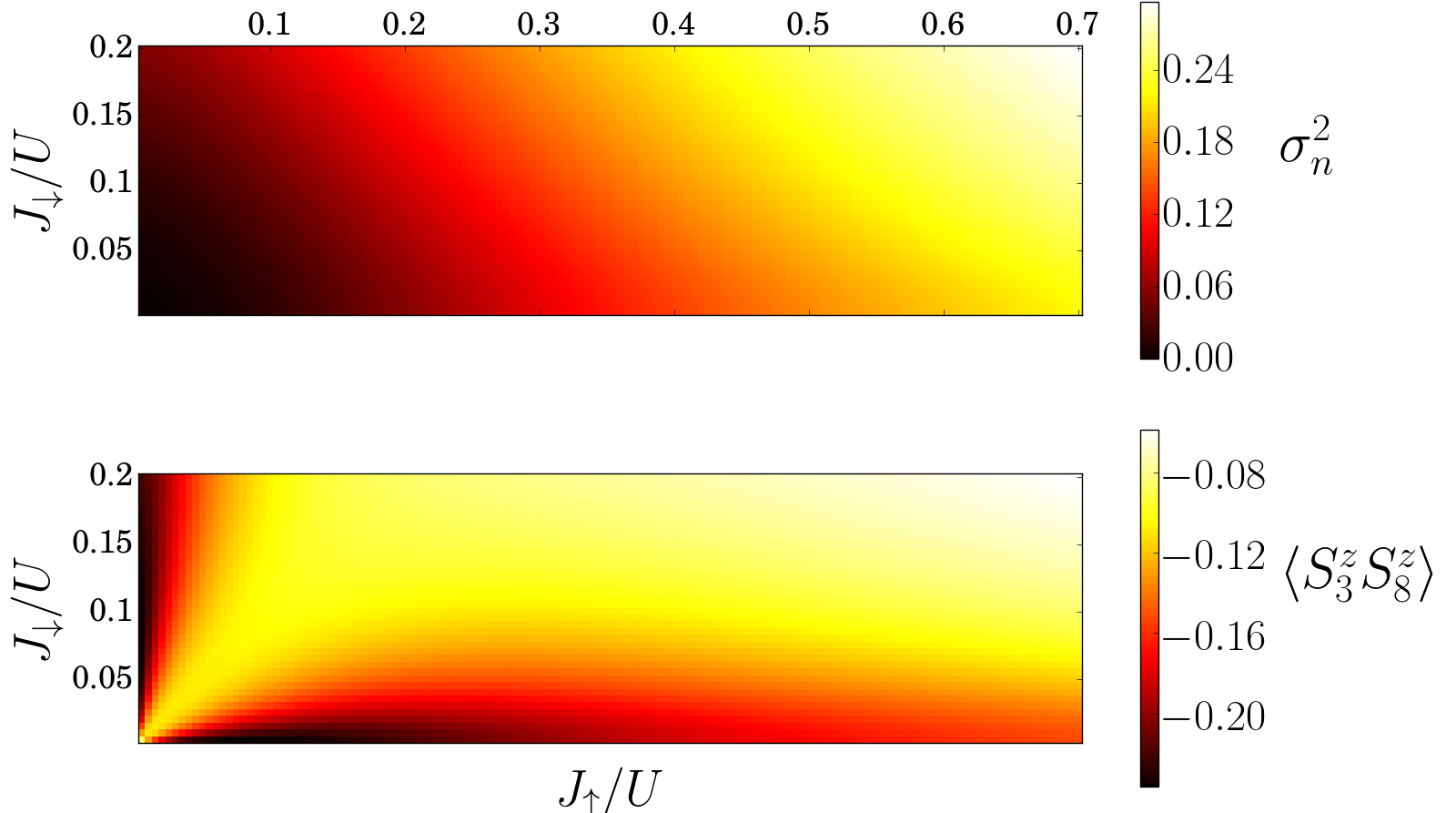}
\end{subfigure}}
\caption{Short and long range spin correlations and on-site number fluctuations in the effective ground state  for Eq. \eqref{eq:anisotropic} with $L=10$ at half-filling. \label{fig:phase}}
\end{figure*}

Figure \ref{fig:phase} shows expectation values computed in the first excited state of $H_A$ for an $L=10$ chain at half-filling as the dimensionless parameters $J_\uparrow/U$ and $J_\downarrow/U$ are varied\footnote{The DMRG calculations were performed with three sweeps; the energy variance (defined as $\langle H^2 - \langle H\rangle^2\rangle$) did not exceed $10^{-7}U^2$ which resulted in a maximum bond dimension of 460. By calculating observables for chains with $L=6$ and $L=16$, we find that finite size effects cause variations in the observables studied by as much as 15\%.}. We plot the first excited state because this is the state which is adiabatically connected to the ferromagnetic initial state of a strong external magnetic field by ramps we study in Sec. \ref{sec:ramps}. In the thermodynamic limit, the first excited state is a microcanonical zero-temperature state.
The left panels show nearest-neighbor spin correlations,  $\langle S^x_5 S^x_6\rangle$ on the top and $\langle S^z_5 S^z_6\rangle$ on the bottom. Both are negative, indicating that the system exhibits antiferromagnetism. Along the diagonal $J_\uparrow = J_\downarrow$, the spin-isotropic model is recovered and the correlators are equal, $\langle S^x_5 S^x_6\rangle = \langle S^z_5 S^z_6\rangle$. In the Ising limit when either $J_\sigma \rightarrow 0$, $\langle S^z_5 S^z_6\rangle$ reach their maximum values.

The bottom right shows longer range correlations along the $z$ axis. These correlations are on the same order as the nearest neighbor correlations, which is indicative of long range magnetic order of the model in the regimes studied. Finally, the top right shows $\sigma_n ^2$ which increases monotonically with $J_\sigma/U$.

\section{Adiabatic Ramps \label{sec:ramps}}

To prepare low-temperature states of $H_A$, we apply a large transverse magnetic field $h(t)$ and initialize the system in the fully polarized band insulating state that is the ground state of this Hamiltonian. We then adiabatically ramp the field to zero. The Hamiltonian describing this setup is
\begin{equation}
H(t)=H_\mathrm{A}-h(t) \sum_i S_i ^x.
\label{eq:ramp}
\end{equation}
To keep the dynamics as adiabatic as possible, the magnetic field should be ramped slowly when the energy gap is small. However, it is not feasible to compute or measure the energy level diagram before each ramp in an experiment, so we consider an exponential ramp,
\begin{equation}
h(t)=h_0 \exp(-t/\tau)-h_0\exp(-t_f/\tau),
\label{eq:ramp}
\end{equation}
where $t_f$ is the final time and $h_0$ characterizes the initial field strength. We shift the exponential ramp shape so that $h(t_f)=0$. In the limit $\tau\rightarrow \infty$, the ramp is fully adiabatic. We choose an exponential ramp shape because it is a convenient choice that reflects the fact that the energy gap tends to be smallest at low $h$. More optimal ramps could be engineered though techniques from quantum control. Even though our exponential ramps are not optimized, we show that they can  still be effective at reaching low-temperature states.

\begin{figure*}[ht!]
\centering
\begin{subfigure}[]{0.49\textwidth}
\centering
\includegraphics[width=.98\linewidth]{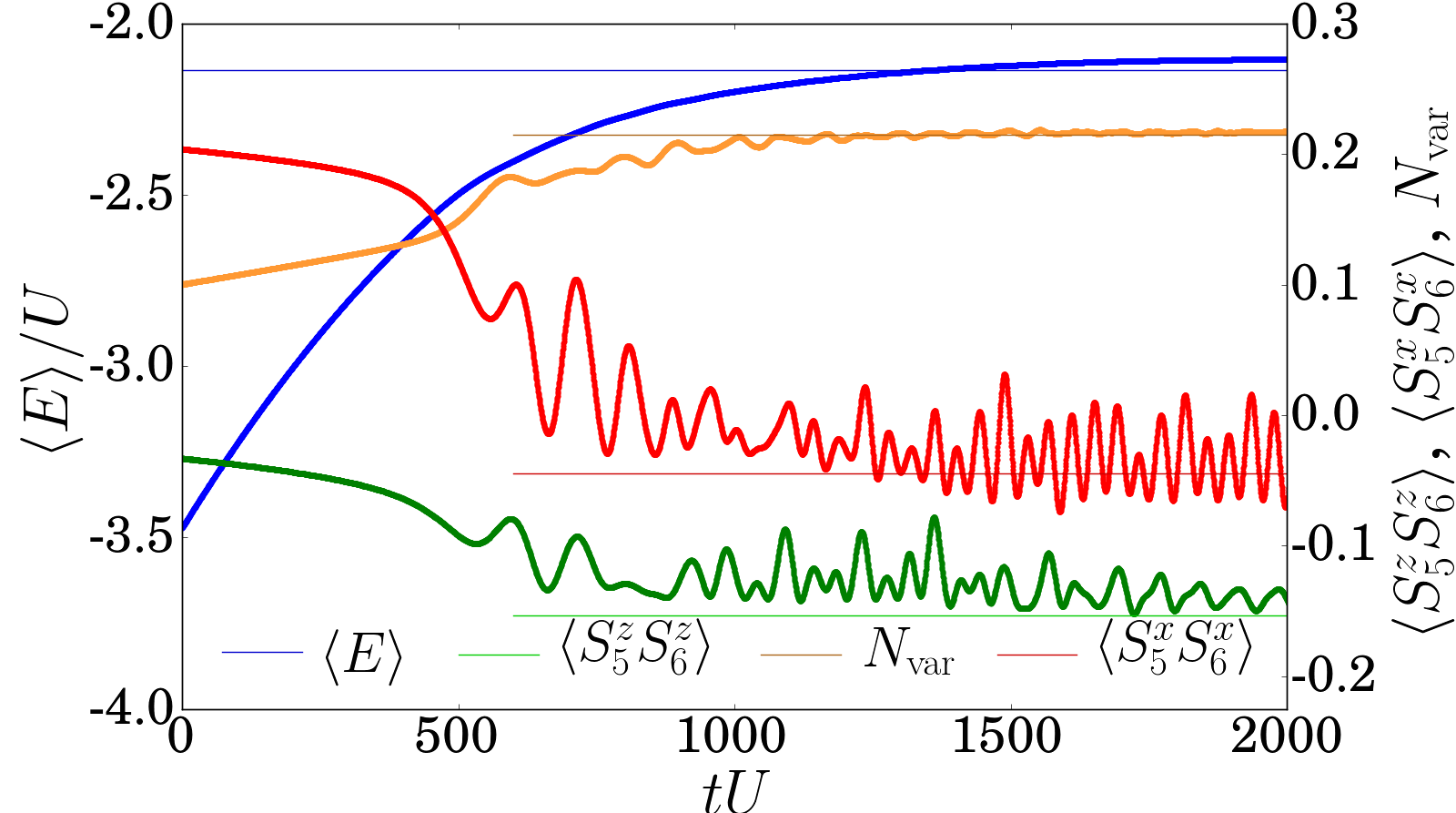}
\end{subfigure}
~
\begin{subfigure}[]{0.49\textwidth}
\centering
\includegraphics[width=.98\linewidth]{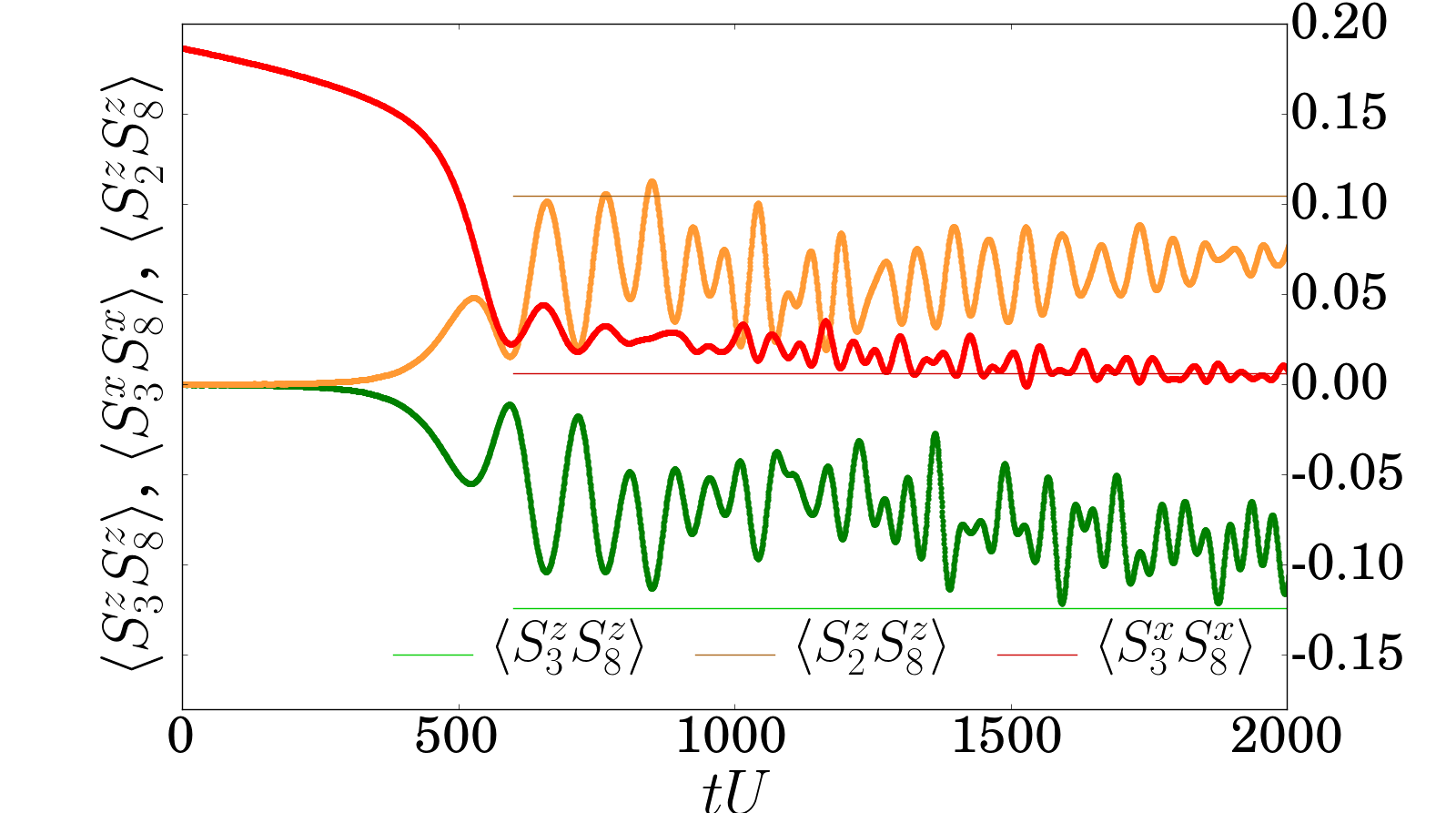}
\end{subfigure}

\begin{subfigure}[]{0.49\textwidth}
\centering
\includegraphics[width=.98\linewidth]{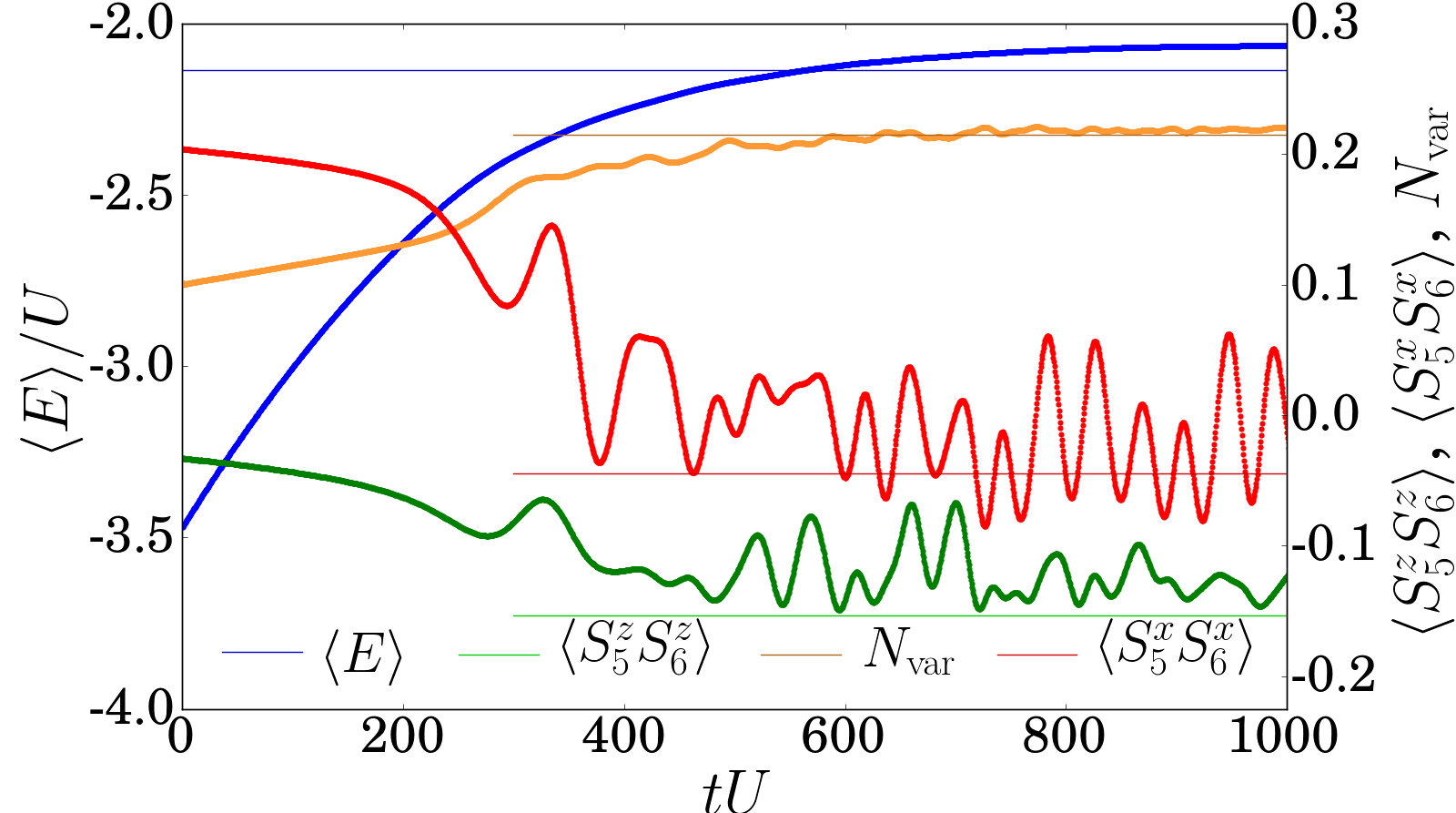}
\end{subfigure}
~
\begin{subfigure}[]{0.49\textwidth}
\centering
\includegraphics[width=.98\linewidth]{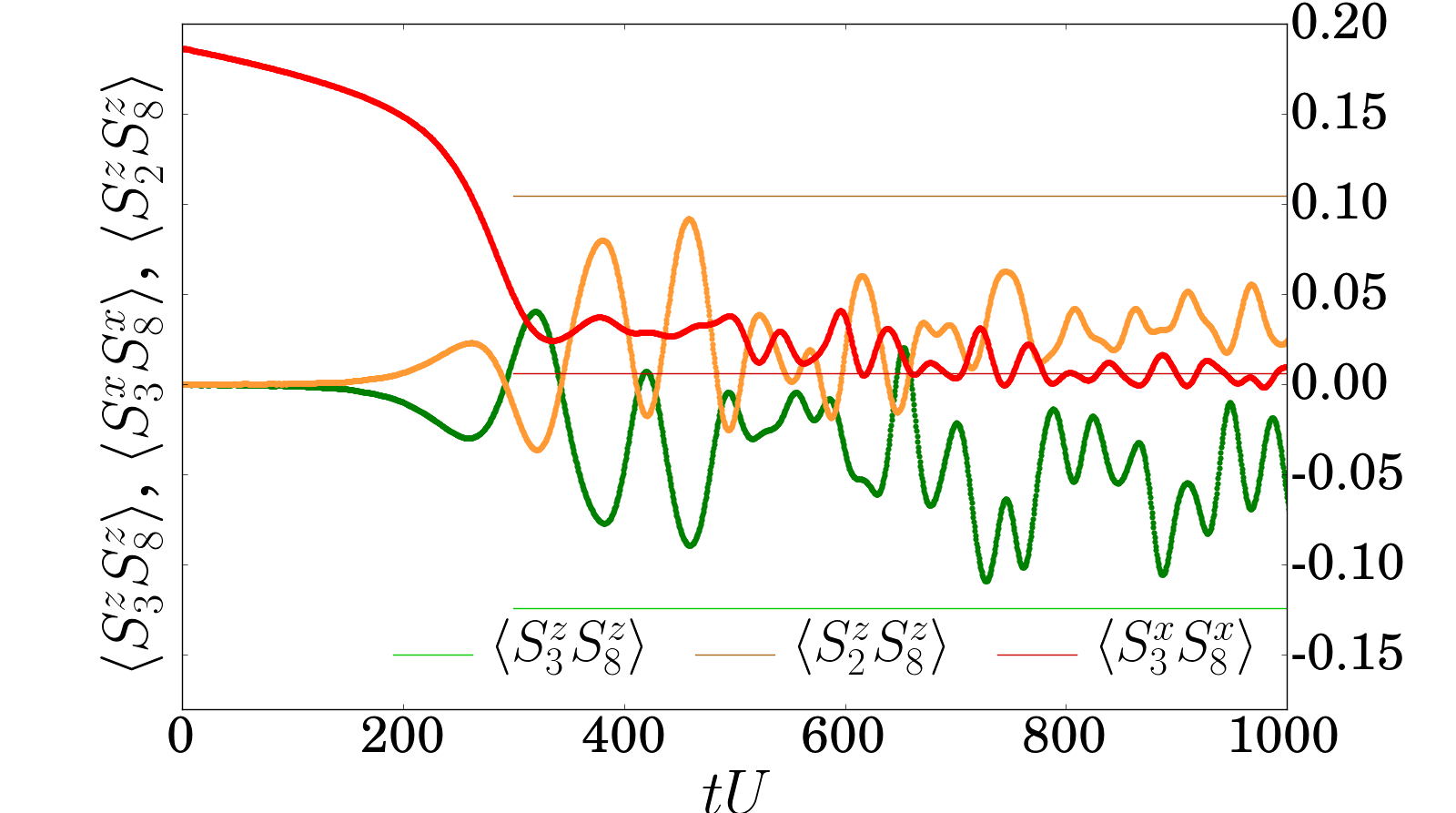}
\end{subfigure}
\caption{Various observables after a ramp of $H(t)$ with $t_\uparrow=0.6U$, $J_\downarrow=0.05U$, $L=10$ at half-filling. In the top panels, $\tau=1000/U$, In the bottom panels $\tau=500/U$. Thin horizontal lines indicate effective ground state observables and thick lines indicate observable dynamics during the ramp. In the top ramp, nearest-neighbor correlations oscillate about a mean value that converges to within 25\% of the effective GS value. In the bottom ramp the mean value of all nearest-neighbor correlations converges within 50\% of the effective GS value In all cases, significant nonzero correlations occur between well-separated sites. \label{overflow}}
\label{fig:ramp}
\end{figure*}

In an experiment, one would take $h_0\gg \{U,J\}$ so that the initial Hamiltonian is dominated by the magnetic field, and the initial state is fully polarized along the $x$-axis, aligned with the transverse field. Such a state can be easily prepared in the lab by rotating the spins into the desired direction with a $\pi/2$ Rabi pulse. In our simulations we assume that the initial state has exactly zero entropy. In experiment, the initial band insulator will have nonzero entropy, as we discuss in Sec. \ref{sec:experiment}. The transverse field can be applied via an oscillating electromagnetic field on resonant with the transition between the states used for the $\left|\uparrow\right >$ and $\left |\downarrow \right>$ states of the model. If these are hyperfine states, a rotating field or microwave field can be used; if these are different electronic states, a laser can be used. 

To reduce computational times, our ramps take $h_0=0.3U$, much weaker than is needed for $h$ to dominate $H_A$, and we ramp according to Eq. \eqref{eq:ramp} with $t_f=2\tau$. That is, we start our calculations after much of the ramp starting from $h\gg \{U,J\}$ has occurred. As such, our initial condition already differs somewhat from the $h\rightarrow \infty$ ferromagnetic state $\left |\rightarrow\rightarrow\rightarrow\cdots \right >$ that is easy to prepare experimentally.
An experiment would start from a much larger $h$ where $\left |\rightarrow \rightarrow \rightarrow \cdots \right >$ is very close to the ground state and ramp down from there. Our procedure is simply for numerical convenience: although it's computationally expensive to simulate the dynamics for the very large $h$, experiments can ramp down quite quickly over this regime since the energy gap between the ground state and any excited state remains large when $h>0.3U$. Consequently, the adiabatic approximation is maintained even for fast ramps in this regime.

Since $H(t)$ lacks the $SU(2)$ symmetry of the original Fermi-Hubbard model due to the spin-dependent tunneling, the ferromagnetic ground state of $H(0)$ is adiabatically connected to a low-energy antiferromagnetic eigenstate of $H(t_f)=H_\mathrm{A}$. However, this eigenstate is not the ground state of $H_\mathrm{A}$: one energy level crossing still occurs during the $h$ rampdown in our model, so that the ferromagnetic initial state is adiabatically connected to the first excited state of $H_A$ rather than the ground state.

This level crossing originates from an additional symmetry of $H(t)$. The Hamiltonian is invariant under a combined lattice reflection and magnetic reflection about the $xy$-plane,
\begin{equation}
c_{i,\sigma}\mapsto c_{n-i+1,-\sigma}.
\end{equation}
This discrete symmetry persists even when $J_\uparrow\neq J_\downarrow$, and it prevents the $t=0$ ground state from connecting to the ground state of $H_\mathrm{A}$.

This is not a problem for reaching zero-temperature states, for two reasons. First, this symmetry is easily broken by a small perturbation on a single site in the lattice. After applying such a perturbation, the initial state is adiabatically connected to the true ground state of the model. Second, in the thermodynamic limit ($L\to \infty$), the difference between any observables evaluated in the ground state and observables evaluated in the first excited state (which we call the ``effective ground state") becomes negligible. This intuitive fact follows, for example, from the eigenstate thermalization hypothesis \cite{ETH}, which states that local observables in an eigenstate are the same as in the canonical ensemble with the same energy density. 

The top panels of Figure \ref{fig:ramp} show the dynamics during a ramp with $J_\uparrow=0.6U$, $J_\downarrow=0.05U$, and $\tau=1000/U$ in an an $L=10$ chain, obtained through DMRG simulations, and demonstrate that observables and correlations out to substantial distances reach values close to the effective ground state\footnote{The DMRG calculations for all ramps studied were performed with a timestep $\mathrm{d}t=0.5/U$, a maximum of 20 sweeps per timestep; the discarded weight did not exceed $10^{-10}$ at each timestep, resulting in a maximum bond dimension of 240.}. In Sec. \ref{sec:experiment}, we show that the parameters used are reasonable for a cold atoms experiment. The left panel shows that energy and $\sigma_n ^2$ converge closely to the effective ground state values after about $3\tau$. Nearest neighbor correlations $\langle S^z _5 S^z _6 \rangle$ and $\langle S^x _5 S^x _6 \rangle$ exhibit substantial oscillations about a mean value that converges near the effective ground state value. However, oscillations remain at all times studied. The right panel shows longer range correlations $\langle S^z _3 S^z _8 \rangle$, $\langle S^x _3 S^x _8 \rangle$, and $\langle S^z _2 S^z _8 \rangle$. The mean value of correlations for both of these $z$ correlations converges to a value which is similar to, but abut 20\% smaller in absolute value than the effective ground state value. $\langle S^x _3 S^x _8 \rangle$ converges closely to its effective ground state value, which is very nearly zero. The agreement of short range and long range spin correlations, energy, and $\sigma_n ^2$ exhibited after the ramp with their effective ground state values each indicate that the ramp is successful in preparing a low-temperature state of the model.

\begin{figure}[h]

\centering
\includegraphics[width=0.48\textwidth]{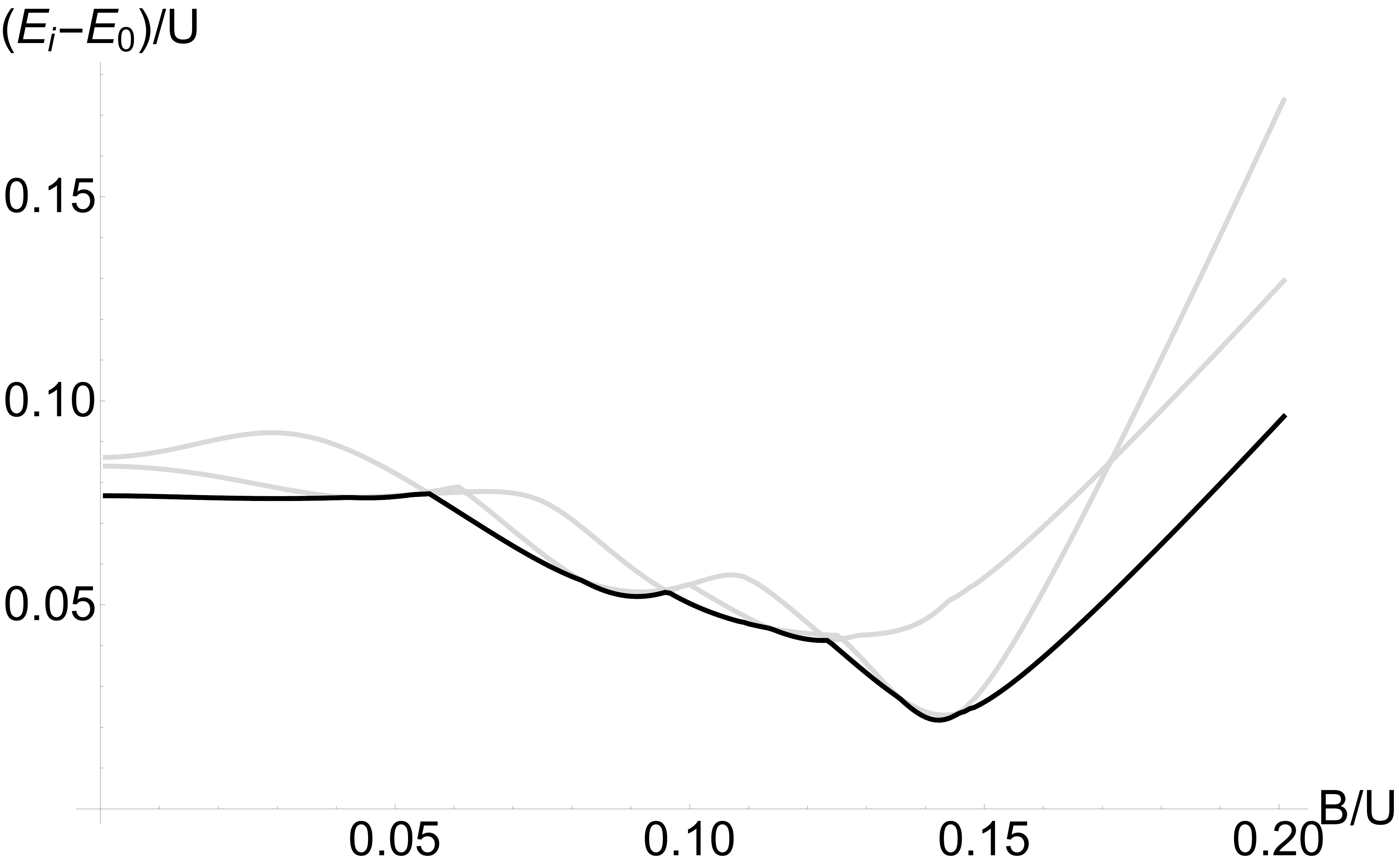}
\caption{Energy gap between excited states and effective ground state of $H_A$ on an $L=10$ chain at half-filling with $J_\uparrow = 0.6U$, $J_\downarrow = 0.5U$ as a function of $h$. Dark line shows energy gap between the effective GS and lowest excited state. \label{overflow}}
\label{fig:levels}
\end{figure}

The oscillations exhibited in all observables except the energy can be understood as an indication that the system is driven into a superposition of different eigenstates as the ramp progresses, i.e. the ramp is not completely adiabatic. The system is excited when the energy gap between the effective ground state and excited states is small. The energy level diagram in Figure \ref{fig:levels} shows that for $J_\uparrow = 0.6 U$, $J_\downarrow = 0.05 U$ in an $L=10$ lattice at half-filling, the energy gap is smallest near $h=0.15U$. This agrees with Figure \ref{fig:ramp}, which shows that the system develops oscillations in its observables when $h$ is lowered through this value (which happens in our ramps around $t=0.5\tau$).

The bottom panels of Figure \ref{fig:ramp} show that even in a shorter ramp, with $\tau=500/U$, all observables except the longest range $S^z$ correlations shown here are close to their effective ground state values. Although all observables deviate more from their effective ground state values than for the $\tau=1000/U$ ramp, the nearest neighbor correlations are within $5\%$ of their ground state values and even the $S^z$ correlations with separations of 5-sites and 6-sites grow to substantial values, about  half of what they achieve in the effective ground state. 

\begin{figure}[h]
\centering
\includegraphics[width=0.48\textwidth]{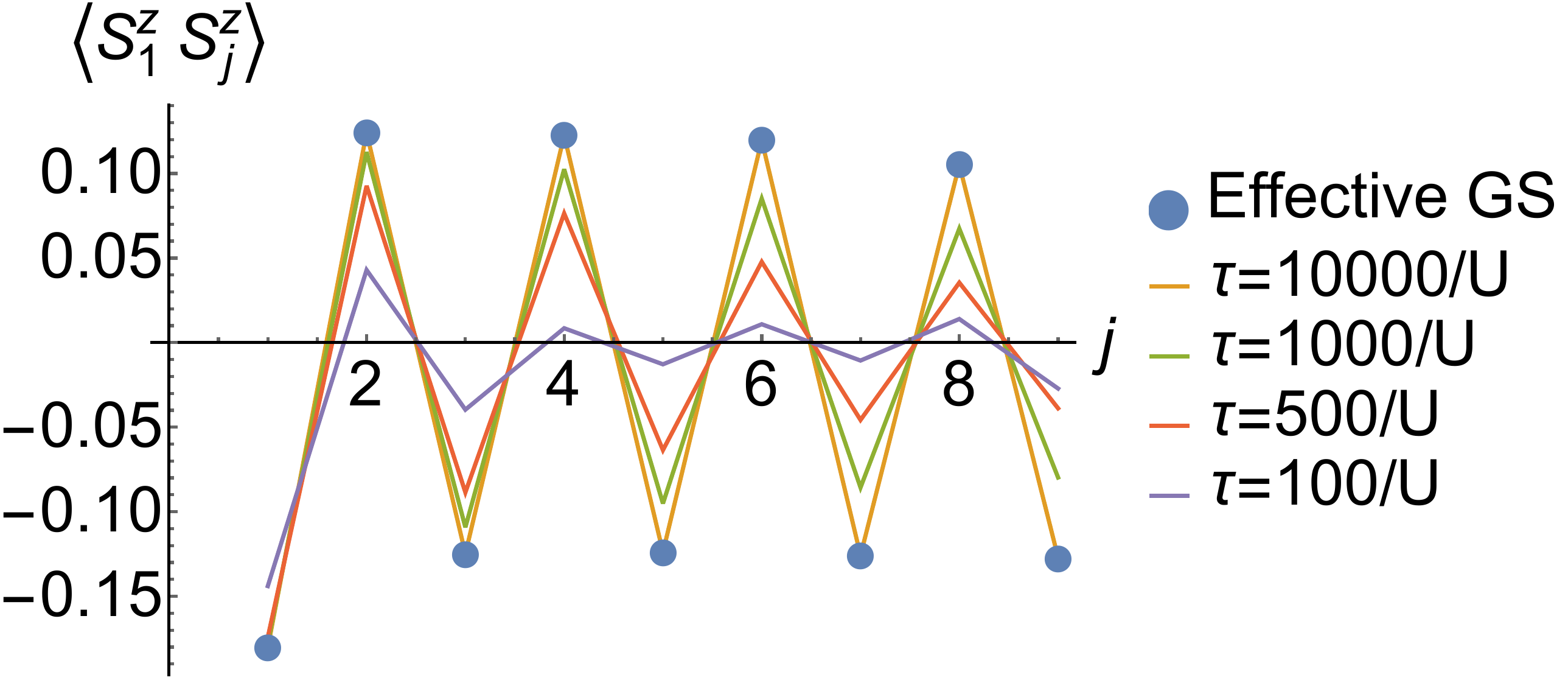}
\caption{Spin correlations along $z$-axis as a function of distance. Plot shows $\langle S_1 ^z S_j ^z\rangle$ vs. $j$, where $j$ indexes the lattice site for ramps with different time constants $\tau$ on an $L=10$ lattice at half-filling with $J_\uparrow = 0.6U$, $J_\downarrow=0.05U$.}
\label{fig:distance}
\end{figure}

The observation that when starting from an uncorrelated state, correlations take more time to develop for sites that are farther apart is a natural consequence of the finite propagation velocity of information, i.e. the ``light-cone" \cite{LiebRobinson}. The rate of this growth can be formally captured by Lieb-Robinson bounds for information propagation. Hence, we expect that the time $\tau$ of the ramp will limit the range of correlations; larger $\tau$ will allow longer-range correlations to develop.  This is borne out by the findings shown in Figure \ref{fig:distance}. The effective ground state value and the final mean values of $\langle S^z _1 S^z _j \rangle$ for $j=2,\ldots,9$ are plotted for ramps of various time constants in a system with $J_\uparrow = 0.6U$ and $J_\downarrow = 0.05 U$. The mean values of observables are calculated by averaging over an interval of $200/U$ after $t_f$. This interval comparable to the timescale of the shortest ramps we study. For the longest ramp, with a time constant of $\tau=10,000/U$, all correlations differ from the effective ground state values by less than $1\%$. As the ramp time $\tau$ is reduced, correlations deviate more from their effective ground state values, with longer range correlations deviating most strongly.

\begin{table*}[ht!]
\centering
\caption{Final state expectation values for adiabatic ramps at various $J_\uparrow/U,J_\downarrow/U,$ and $\tau$. For all ramps, $h_0=0.3U$, $t_f=2\tau$. Observables are calculated by averaging over an interval of $200/U$ after $t_f$. For $\tau\sim 10,000$, all observables converge within two significant digits. For shorter ramps, short-range correlation converge better than long-range correlations. Convergence is best for highly anisotropic $J_\uparrow$ vs. $J_\downarrow$. }
\label{my-label}
\begin{tabular}{cclrrrrr}
\multicolumn{1}{l}{\multirow{2}{*}{$J_\uparrow / U$}} & \multicolumn{1}{l}{\multirow{2}{*}{$J_\downarrow / U$}} &                & \multicolumn{1}{l}{\multirow{2}{*}{$\langle E\rangle / U$}} & \multicolumn{1}{l}{\multirow{2}{*}{$\left < S^z _5 S^z _6 \right >\times 10$}} & \multicolumn{1}{l}{\multirow{2}{*}{$\left < S^z _3 S^z _8 \right >\times 10$}} & \multicolumn{1}{l}{\multirow{2}{*}{$\left < S^x _5 S^x _6 \right >\times 100$}} & \multicolumn{1}{l}{\multirow{2}{*}{$N_\mathrm{var}\times 10$}} \\
\multicolumn{1}{l}{}                                  & \multicolumn{1}{l}{}                                    &                & \multicolumn{1}{l}{}                                        & \multicolumn{1}{l}{}                                                           & \multicolumn{1}{l}{}                                                           & \multicolumn{1}{l}{}                                                            & \multicolumn{1}{l}{}                                           \\ \hline
\multirow{5}{*}{0.6}                                  & \multirow{5}{*}{0.05}                                   & $\tau=100/U$   & -1.9209                                                     & -1.047                                                                         & -0.148                                                                         & -1.16                                                                           & 2.232                                                          \\
                                                      &                                                         & $\tau=500/U$   & -2.0643                                                     & -1.313                                                                         & -0.617                                                                         & -2.59                                                                           & 2.192                                                          \\
                                                      &                                                         & $\tau=1000/U$  & -2.1050                                                     & -1.408                                                                         & -0.915                                                                         & -3.40                                                                           & 2.172                                                          \\
                                                      &                                                         & $\tau=10000/U$ & -2.1346                                                     & -1.526                                                                         & -1.235                                                                         & -4.49                                                                           & 2.162                                                          \\
                                                      &                                                         & GS             & -2.1349                                                     & -1.531                                                                         & -1.242                                                                         & -4.49                                                                           & 2.149                                                          \\ \hline
\multirow{2}{*}{0.5}                                  & \multirow{2}{*}{0.05}                                   & $\tau=1000/U$  & -1.6034                                                     & -1.490                                                                         & -0.936                                                                         & -3.82                                                                           & 1.864                                                          \\
                                                      &                                                         & GS             & -1.6370                                                     & -1.608                                                                         & -1.329                                                                         & -4.70                                                                           & 1.860                                                          \\ \hline
\multirow{5}{*}{0.3}                                  & \multirow{5}{*}{0.05}                                   & $\tau=100/U$   & -0.5133                                                     & -0.085                                                                         & -0.087                                                                         & 2.38                                                                            & 0.875                                                          \\
                                                      &                                                         & $\tau=500/U$   & -0.6508                                                     & -1.385                                                                         & -0.064                                                                         & -4.82                                                                           & 1.029                                                          \\
                                                      &                                                         & $\tau=1000/U$  & -0.6884                                                     & -1.514                                                                         & -0.584                                                                         & -2.56                                                                           & 1.064                                                          \\
                                                      &                                                         & $\tau=10000/U$ & -0.7462                                                     & -1.746                                                                         & 1.469                                                                          & -5.48                                                                           & 1.117                                                          \\
                                                      &                                                         & GS             & -0.7467                                                     & -1.758                                                                         & -1.495                                                                         & -5.48                                                                           & 1.115                                                          \\ \hline
\multirow{2}{*}{0.6}                                  & \multirow{2}{*}{0.07}                                   & $\tau=1000/U$  & -2.1211                                                     & -1.128                                                                         & -0.528                                                                         & -2.09                                                                           & 2.230                                                          \\
                                                      &                                                         & GS             & -2.2013                                                     & -1.391                                                                         & -1.086                                                                         & -4.66                                                                           & 2.249                                                          \\ \hline
\multirow{2}{*}{0.5}                                  & \multirow{2}{*}{0.07}                                   & $\tau=1000/U$  & -1.6209                                                     & -1.357                                                                         & -0.582                                                                         & -3.22                                                                           & 1.974                                                          \\
                                                      &                                                         & GS             & -1.7002                                                     & -1.458                                                                         & -1.165                                                                         & -4.85                                                                           & 1.957                                                          \\ \hline
\multirow{2}{*}{0.3}                                  & \multirow{2}{*}{0.07}                                   & $\tau=1000/U$  & -0.6915                                                     & -1.004                                                                         & -0.062                                                                         & -2.94                                                                           & 1.091                                                          \\
                                                      &                                                         & GS             & -0.8013                                                     & -1.553                                                                         & -1.297                                                                         & -5.49                                                                           & 1.203                                                         
\end{tabular}
\end{table*}

The convergence of observables to their effective ground state values and the development of long range correlations are not particular to the parameters chosen for the ramps in Figure \ref{fig:ramp}. In Table 1, observables are computed after the ramping procedure is completed in an $L=10$ lattice at half-filling for numerous choices of the dimensionless parameters $J_\uparrow/U$ and $J_\downarrow/U$, demonstrating that the effective ground state with correlations between well-separated sites can be prepared in a robust region of parameter space. The parameters are all chosen in the regime $U/J\sim 10$, as many experiments operate in this regime where antiferromagnetic correlations are strongest \cite{Greiner}.

\section{Experimental Realizability \label{sec:experiment}}

Several experimental methods have been developed to implement state-dependent lattices needed for our demagnetization protocol. One straightforward procedure is to employ a dual species experiment. Each atomic species experiences a different trap depth and hence different tunneling rates \cite{DualSpecies,Thywissen,DualExp}. However, we note that this is insufficient for our demagnetization scheme, because the different species will experience different effects from the transverse field.

To implement both the spin-dependent lattice and the transverse field, there are at least three possible techniques. 
One can use alkaline earth atoms, using the long-lived $^3P_0$ clock state and ground electronic state as the two levels \cite{alkaline}. These states typically experience different optical lattice depths as they have different AC polarizabilities at most wavelengths.
Alternatively, a rapidly oscillating transverse field gradient can be applied and used to tune the respective tunneling energies of different hyperfine states \cite{oscillate}. If the frequency of the oscillating field gradient is chosen carefully, the associated heating processes can be suppressed \cite{Eckardt} Finally, a spin-dependent lattice can be implemented by detuning near the hyperfine splitting \cite{Bloch,Porto}. In all cases, when these lattices are applied to ultracold fermionic atoms, a Hubbard model with spin-dependent tunneling rates is realized.

An assumption of our calculations is that the initial state is in the ground state of the Hamiltonian with $h_0=0.3U$. This is an idealization: in experiment it is not possible to ramp from a zero entropy spin-polarized band insulator to the $h_0=0.3$ ground state perfectly adiabatically, though since the energy gap is large, the entropy introduced by this process will not be significant compared to the heating caused during the ramp from $h_0$ to zero. 

More importantly, it is also not possible to prepare a spin-polarized band insulator at zero entropy. Since adiabatic demagnetization cannot remove entropy from a system, any initial entropy will remain in the final state. For example, Ref. \cite{badband} produces band insulators with entropies on the order of $0.25 k_B$ per particle. In principle, this is low enough to prepare states below the N\'eel temperature if the ramp was perfectly adiabatic. Furthermore, the current state of the art, as presented in Ref. \cite{goodband}, achieves a doublon band insulator with less than $0.02 k_B$ per particle, a negligible amount.

The parameters chosen in the simulations are experimentally realizable with ultracold atoms. The lattice depth and atomic species can be chosen so that the dimensionless parameters $J_\uparrow/U$, $J_\downarrow/U$ are on the order of the ratios shown in Table 1, while the timescales of the ramps remain manageable. For example, with an up-spin lattice depth $V_\uparrow \sim 5 E_R$, a down-spin lattice depth $V_\downarrow \sim 15 E_R$, and an $s$-wave scattering length $a_s\sim 0.02 a$, where $E_R=\hbar^2 \pi^2/2ma$ is the recoil energy, $a$ is the lattice spacing, and $m$ is the mass of the atoms, we obtain the dimensionless parameters $J_\uparrow/U \sim 0.3$, $J_\downarrow/U \sim 0.05$ \cite{jaksch}. At these lattice depths and lattice width, the system is well described by a Hubbard model.

Since our ramps are not optimized, the timescales we currently require are only just within experimental lifetimes of ultracold matter. For lithium in a lattice of wavelength 532 nm, for example, the recoil energy is on the order of $30\,\mathrm{kHz}$. In this case, the onsite interaction is $U \sim 6$ kHz, leading to a time constant for the exponential ramps $\tau=1000/U$ on the order of $200\,\mathrm{ms}$. For our exponential ramps which run for $2\tau$, this is less than typical lifetimes of ultracold matter \cite{emission}.

The ratios for $J_\sigma/U$ studied here have not been optimized for experimental implementation, and the ramp times can likely be substantially decreased by fine tuning these parameters. There is also significant room for optimization beyond the exponential ramps considered here that could bring down the timescales of the ramps. We optimized such a ramp for an $L=4$ lattice, where the methods of Ref. \cite{geodesic} could be easily implemented. With these methods, we designed highly adiabatic ramps which produced low-temperature states in a total time of $100/U$. 

\section{Conclusions \label{sec:Conclusion}}
We propose  a protocol to use adiabatic demagnetization from a readily-prepared spin-polarized band insulator to low-temperature states of the Fermi-Hubbard model with spin-dependent tunnelings in Eq. \eqref{eq:anisotropic}. The spin-dependent tunnelings overcome the impediment to adiabatic demagnetization in the usual Fermi-Hubbard model. This procedure could be the first step of preparing ground states of the usual Fermi-Hubbard model, by following it by adiabatically bringing the parameters to $J_\uparrow=J_\downarrow$. Moreover, similar techniques could be developed to access other low temperature states of spin models, such as the high temperature superconducting phases, and our scheme also works in principle at densities besides half-filling. Thus it could give a route to accessing other behaviors in the Hubbard model, e.g. bad metal and pseudogap behaviors, and superconducting phases.

To quantitatively assess how efficient this protocol is, we numerically calculated the dynamics of observables including spin correlations  under ramps that can realistically be implemented in cold atoms experiments. Our calculations were for one-dimensional systems, where we can employ DMRG. The model's low-energy states exhibit long range antiferromagnetic order, and our adiabatic protocol successfully develops these long range correlations in a robust region of the phase diagram.

Finally, we note that our procedure for generating ground states of the spin-anisotropic model could be adapted to generate ground states of the spin-isotropic Fermi-Hubbard model. In principal the ground state of the spin-isotropic model is adiabatically connected to the effective ground state of $H_A$ by tuning the tunneling strengths. Once the ground state of the anisotropic Hamiltonian is prepared, the tunneling strengths can be ramped to obtain a ground state of the original Hubbard Model. We expect that the timescales required to maintain adiabaticity may be similar to those of the ramps described above and therefore compatible with the timescales available in experiment.

\section*{Acknowledgements}
We thank Randy Hulet and Ian White for discussions. This work was supported in part with funds from the Welch Foundation, Grant No$.$ C-1872. K.R.A.H. thanks the Aspen Center for Physics, which is supported by the National Science Foundation grant PHY-1066293, for its hospitality while part of this work was performed. This work was supported in part by the Data Analysis and Visualization Cyberinfrastructure funded by NSF under grant OCI-0959097 and Rice University.

\bibliographystyle{apsrev4-1}
\bibliography{bibliography.bib}

\end{document}